\documentclass[journal]{IEEEtran}

\usepackage{cite}
\usepackage[pdftex]{graphicx}
  \graphicspath{{pics/}}
  \DeclareGraphicsExtensions{.pdf,.jpeg,.png,.jpg}
\usepackage{url}

\begin{document}

\title{Fog Intelligence for Network Anomaly Detection}

\author{Kai~Yang, Hui~Ma, and Shaoyu~Dou
\thanks{Kai Yang, Hui Ma, and Shaoyu Dou were with the Department of Computer Science, Tongji University, China.
E-mail: kaiyang@tongji.edu.cn}}% <-this % stops a space

\maketitle
\begin{abstract}

Anomalies are common in network system monitoring. When manifested as
network threats to be mitigated, service outages to be prevented, and security risks to be ameliorated, detecting such anomalous network behaviors becomes of great importance. However, the
growing scale and complexity of the mobile communication networks, as well as the
ever-increasing amount and dimensionality of the network surveillance data, make it extremely
difficult to monitor a mobile network and discover abnormal network behaviors.
Recent advances in machine learning allow for obtaining near-optimal solutions
to complicated decision-making problems with many sources of uncertainty that
cannot be accurately characterized by traditional mathematical models. However,
most machine learning algorithms are centralized, which renders them
inapplicable to a large-scale distributed wireless networks with
tens of millions of mobile devices. In this article, we present fog intelligence,
a distributed machine learning architecture that enables intelligent wireless
network management. It preserves the advantage of both edge processing and
centralized cloud computing. In addition, the proposed architecture is
scalable, privacy-preserving, and well suited for intelligent management
of a distributed wireless network.
\end{abstract}

\begin{IEEEkeywords}
Artificial Intelligence, Anomaly Detection, Fog and Edge Computing, Network Intelligence, 5G systems.
\end{IEEEkeywords}

\IEEEpeerreviewmaketitle

\section{Introduction}
% description of wireless networks
With the rapid advancements of modern communication and signal processing technologies,  wireless communications are becoming ubiquitous in our everyday life. Future Fifth Generation (5G) network is envisioned to be heterogeneous, software-defined, and far more complicated than the Fourth-Generation (4G) system. The 5G system may consist of millions of base stations and serve tens of millions mobile users as well as IoT devices \cite{xiao2017millimeter,YangOFDMA2009}. In addition, the traffic conditions and the Radio Frequency (RF) environments can be highly dynamic which pose great challenges against the network operation and management. The network management is particularly challenging for mission-critical applications such as health care and industrial IoT systems which need to meet stringent delay requirements and work without failures\cite{Zhang2018Approaches}.

The operation and management of the Fourth Generation (4G) wireless networks such as anomaly detection and root cause analysis are semi-autonomous or manual, which are time-consuming and costly\cite{yang2017deep}.
The advent of artificial intelligence and machine learning technologies opens up new avenues for intelligent network management that can automatically detect and diagnosis network failures, network intrusion, as well as performance degradation, leading to better user experience with lower operational costs. However, traditional machine learning algorithms are centralized, which render them difficult to be implemented on a distributed wireless networks with numerous nodes. In addition, transmitting vast amount of data from wireless devices to a single server requires large network bandwidth and may cause excessive delays in the processing.

As a remedy, distributed approaches such as edge computing \cite{Shi2016Edge,Du2007AnEK, li2018learning,DRO,Du2009TransactionsPA} has been proposed that seeks to process data close to the locations where the data was generated. Since the data is processed locally it requires less network bandwidth compared to the centralized approach and consequently may reduce the processing cost and delay while improving the responsiveness and scalability. However, the edge devices are typically of limited power and computing resources and thus may not have the capability of carrying out sophisticated machine learning tasks. The fog computing architecture, on the other hand, represents a horizontal architecture that aims to realize a seamless computing service by utilizing resources from both the cloud server and the edge devices. Inspired by the concept of fog computing, we propose a fog-intelligence architecture for smart network management and operation in this paper. Fog intelligence is flexible and aims to realize the network intelligence through distributed machine learning algorithms on a hierarchical architecture composing by both the cloud and edge computing platforms.

The remainder of this paper is organized as follows. Section II introduces basic concepts of intelligent network management of wireless network. Section III provides a brief survey about distributed machine learning models and algorithms, with emphasis on distributed deep learning. In Section IV, we discuss how distributed machine learning algorithms can enable autonomous management of a complex wireless networks. An example of autonomous network anomaly detection and root cause analysis is also presented. We conclude this paper in Section V.

%---------------------------Shaoyu-Dou----------------------------------------------------------
\section{Intelligent Service Quality Management for 5G Networks}

%what is SQM
With the explosive growth of smartphones and other mobile devices,
the network environment of the cellular system is becoming more complex than ever,
and securing quality of communication services in areas with dense mobile users
has become a major challenge for network operators. Such a challenge will become increasingly challenging for the 5G system since it is heterogeneous and more complex than the 4G system. The Service Quality Management(SQM) system is widely used to ensure the Quality of Service for a communication network. However, SQM is traditionally carried out manually or semi-automatic, which is time-consuming and of high cost. The recent advance in machine learning and artificial intelligence opens up new avenues for intelligent SQM. According to EU H2020 SELFNET project \cite{neves2016selfnet}, network intelligence is one of the major innovative aspects of 5G, which enables intelligent SQM for 5G networks.

%SQM defination
An SQM system can be used to assess quality of service(QoS) qualitatively or quantitatively,
or to compare the gap between the quality of service delivered to mobile users
and the expected quality of service.
In particular, an SQM system monitors a set of performance indicators
and generates alerts whenever service delivery is abnormal.
The goal of an SQM system for communication networks can be defined from two different perspectives.
From the perspective of the network operator,
QoS is defined by the performance indicators of the network,
such as network delay, data packet loss rate, network jitter and other performance indicators.
From this perspective, an SQM is considered to ensure the network performance.
From the perspective of mobile users, QoS highly depends on the particular type of service that the network is expected to deliver to a mobile users.
And SQM system needs to be carried out to ensure the QoS of different types of services for a variety of users. The main purpose of SQM is to make service quality consistent with business objectives
and guarantee the Quality of Experience(QoE) of users.

%how to define Qos
%select performance indicators
As the basis of SQM, how to quantify and calculate the QoS is a primary challenge.
Since the experience of user is a subjective feeling,
it is difficult to accurately characterize this feeling using one or a set of indicators.
Traditionally, a set of thresholds for each indicators are used to quantitatively characterize the QoS,
in which the exact thresholds depend on the consensus between the user and the service provider.
The selection of indicators is usually based on a service quality model such as \emph{SERVQUAL}.
\emph{SERVQUAL} is a QoS measurement tool that captures user expectations of services from five dimensions, i.e., reliability, assurance, empathy, responsiveness, and tangibles. This model is suitable for most communication networks, but so far there exists no universal solution to calculate specific service levels for different types of services.

SQM involves a number of tasks, including network anomaly detection and diagnosis, network resource allocation for QoS optimization etc.
The performance of a communication network can be characterized by a collection of network performance indicators, such as dropped-call rate.
Anomalies in the network performance indicators often represent abnormal conditions of a communication network that requires investigation and diagnosis.
Anomaly detection can be cast as an unsupervised learning problem and has recently received significant research interest.
However, in practice, network operators are usually interested in only a subset of network anomalies that reflect the network performance degradation.
Therefore it remains a great challenge to design anomaly detection algorithm by tailoring general machine learning models to the specific problem under investigation.

%----------------------------------------Hui-Ma-------------------------------------------
\section{Fog-Intelligence for Network Anomaly Detection}

\subsection{Fog-Intelligence Architecture}

Machine learning (ML) based network management usually involves three stages, namely model selection, model building and model deployment. In the first step, an ML model is selected according to the particular network management problem under investigation. Once an ML model is selected, a set of training data is used to calculate the hyperparameters of this model by the ML algorithm. Then the obtained ML model will be deployed to devices for network management. built by applying the training algorithm a set of training data. Please note that the experience from domain expert can also be taken into account in the model building process through methods such as active learning. Machine learning algorithms can be broadly divided into three groups, namely the supervised machine learning, unsupervised machine learning, and reinforcement learning. Supervised ML methods include linear regression, random forest, Support Vector Machine, and deep neural network. Supervised machine learning methods require certain amount of labels provided by human experts. On the contrary, unsupervised machine learning algorithm is capable of drawing inference from the input data without the labelling process, and thus is particularly suitable for problems in which the human labels are difficult to obtain, such as anomaly detection. The anomaly detection aims to find out the anomalous behaviors. These unusual patterns are often regarded as anomalies or outliers. With the sharp rising of ubiquitous and intricate Internet, anomaly detection has been one of the hot topic in the area of network security which yields lots of attention. For instance, the intrusion detection systems (IDSs) provides host or network with security mechanism avoiding the unauthorized entry and typical attacks.

ML based network management system can be realized on a centralized system, which means both the model building and inference drawing are conducted on a single location such as data center or cloud site, as shown in Figure 1. The centralized approach allows for easy maintenance as well as better data privacy, system security and resilience. A centralized site often has large amount of computational power to process the training data, build the ML model, and then draw the inference from the incoming data from entire network. Such an approach also allows the network operator to have a global view about the network and can reduce the cost because of the economics of scale  \cite{calo2017edge}. However, the centralized approach also has some disadvantages. Since the data needs to be collected from the probes and sensors of the entire network to a central location which requires a large amount of network bandwidth. The data transmission is particularly challenging for 5G network due to its increased complexity and heterogeneity. In addition, if the inference is also drawn on the central location, the control signal needs to be sent back to end devices for network management, which may incur significant network delay depending on the network traffic conditions. The transmission delay can be a serious problem for mission-critical applications such as autonomous driving.
\begin{center}
\includegraphics[width=3in]{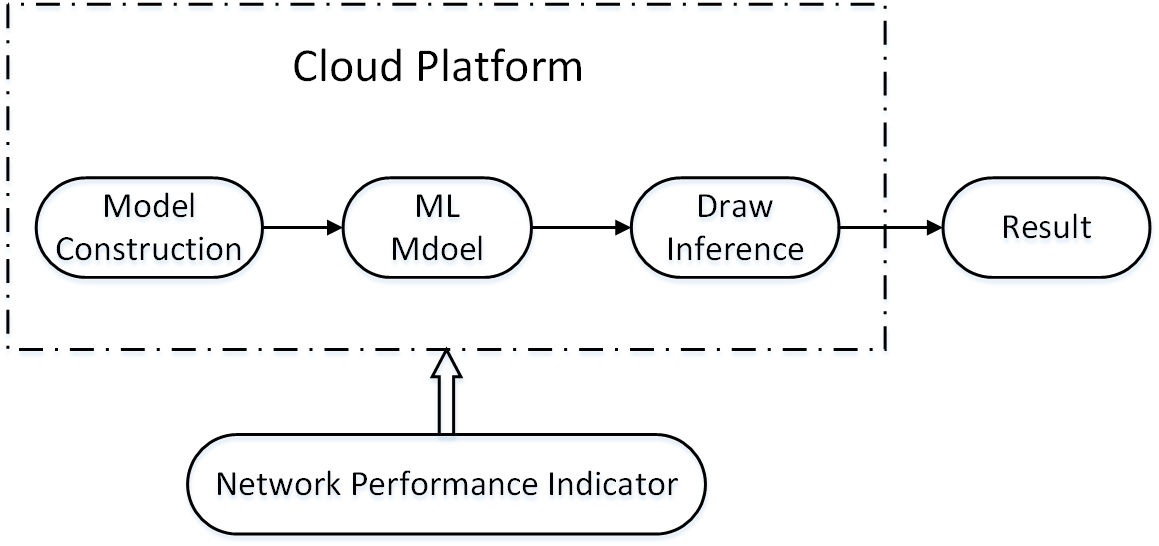}\\
{\ Figure 1.}  Centralized AI for Network Management
\end{center}

ML based network management system can also be implemented on an edge computing architecture. In this edge-intelligence approach, the model building is conducted on a centralized place such as cloud computing platform. However, as opposed to the centralized approach, the inference is drawn at the edge device instead of the centralized place. Since the inference task is carried out at the edge device, the edge intelligence can significantly reduce the transmission delay and enable real-time processing for applications such as autonomous driving. However, since the 5G network is very complex, transmitting the training data from the entire network to a single place requires huge amount of network bandwidth and pose a great challenge against practical implementations.

5G network is heterogeneous and consists of a wide variety of network elements such as cellular base stations and radio network controller with different processing capabilities. We therefore propose a fog-intelligence architecture for intelligent network management in this paper. As shown in Figure 2,  fog-intelligence seeks to realize network intelligence through distributed machine learning algorithms over wireless networks. Different from the edge-intelligence architecture in which the machine learning model is built on a centralized server and then disseminated to edge devices, the machine learning model can be trained in a distributed fashion over multiple fog nodes.  However, please note that the fog-intelligence architecture is flexible and can harness the power of both cloud and edge computing platforms. In addition, the network intelligence can be realized anywhere along the cloud-to-edge continuum, depending on the applications, the privacy, and data protection requirements. For example, in order to guarantee QoS of users, a centralized server can be used to collect surveillance data such as Call Detailed Record (CDR) data from all cells and build an AI model based on the vast amount of collected data. The trained model is then disseminated to each base station to gain insight into the conditions of each cell. While this approach allows the central server to collect data from the entire network which offers a large amount of data for training the AI model, it has its own disadvantages. In particular, the user-level CDR data contains a lot of private information of the user and consequently sending such information from all users to a central server may cause concerns of the data privacy. In addition, sending all the monitoring data to a central server will inevitably cost a lot of network bandwidth and may leads to excessive delay in the processing. In order to preserve the data privacy of each user, we can aggregate the CDR data at the cell level, and then send the aggregated data to a local proximity such as a server close to radio network controller (RNC) for model training. We next discuss a set of distributed machine learning frameworks that can be used to build a fog-intelligence system.
\begin{center}
\includegraphics[width=3in]{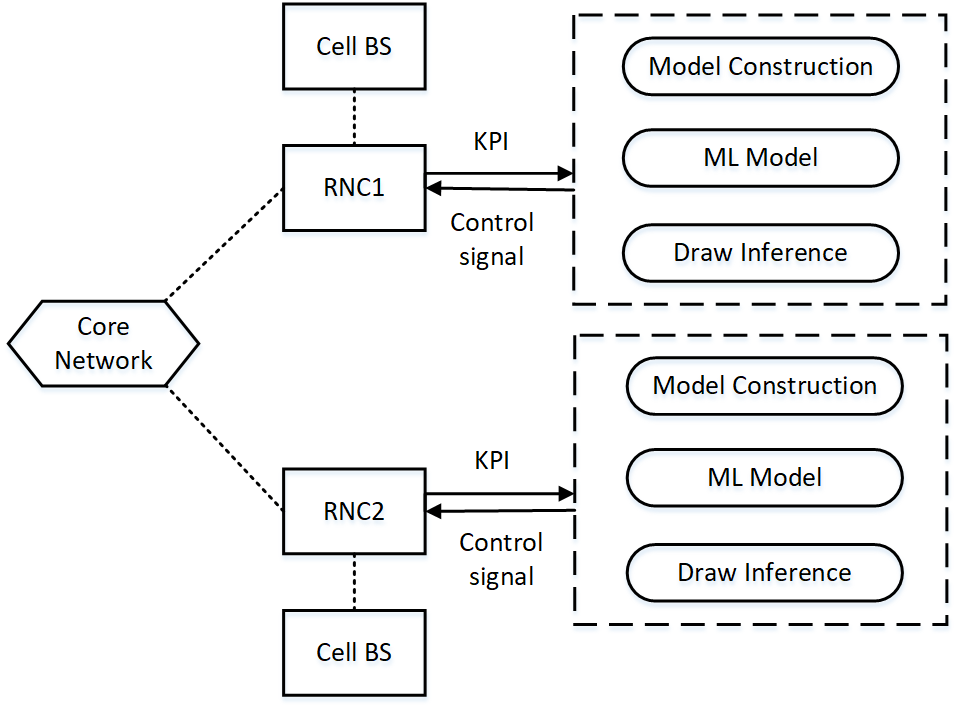}\\
{\ Figure 2.}  Fog-Intelligence for Network Management
\end{center}

\subsection{Distributed Machine Learning Platforms for Fog-Intelligence}

One distinct advantage of the fog intelligence is that both the model training and inference drawing can be carried out in a distributed manner over the entire network, which improves the scalability and responsiveness of the system. We next discuss a number of distributed ML platforms and algorithms that can be used to build an intelligent network management system to detect anomalous network behaviors.

Model parallelism and data parallelism are two different approaches to realize an ML algorithms in a distributed fashion. In the model parallelism approach, the same data is sent to multiple machines and each machine is responsible for calculating a portion of the parameters. These machines then exchange the calculated parameters with each other to come up with all the parameters of the model. If properly designed, the model parallelism approach can significantly reduce the training time. In the data parallelism approach, the training datasets are divided into multiple portions, which are used to calculate all model parameters in parallel. The obtained parameters are then exchanged with each other to obtain the final estimation of the parameters. These two approaches can also be combined to further improve the algorithm efficiency. There are a number of distributed ML platforms based on model parallelism, data parallelism, or a combination of them, including Petuum\cite{xing2015petuum}, Angel\cite{jiang2017angel},Spark\cite{hosseini2018robust} and Distributed TensorFlow\cite{abadi2016tensorflow}. We briefly discuss Spark and Distributed TensorFlow as follows.

Apache spark provides a comprehensive, easy-to-use framework for managing distributed big data processing of datasets and data sources with different properties (text data, chart data, etc.). Because of the high efficiency and speed, Spark has been widely used in many companies around the world. Within the framework of Spark, the Spark's MLlib contains a variety of functional machine learning algorithms. Besides, the Apache spark achieves the high performance for both batch and streaming data. Nevertheless, Spark has limited capacity in dealing with high-dimensional models.

DistBelief is an alternative approach for distributed machine learning. It works well with model parallelism, but in terms of data parallelism, it only supports optimization on a single objective. As a remedy, TensorFlow has been proposed. TensorFlow\cite{abadi2016tensorflow} is a platform for machine learning algorithms which can be implemented on different types of hardware ranging from mobile devices to large-scale computing clusters. TensorFlow is available on different operation systems including Windows, Linux, macOS, as well as mobile computing platforms including Android and iOS. TensorFlow can either work in a single machine or runs in distributed environment over different devices, which is called the distributed TensorFlow \cite{abadi2016tensorflow}. Distributed TensorFlow has been applied to a wide range of areas including robotics, computer vision, object detection, speech recognition, natural language processing, etc.

TensorFlow Lite is designed to use TensorFlow for both embedded and mobile devices. It is lightweight and can be deployed on mobile computing platforms including Android and iOS systems. In addition, TensorFlow Lite can support a number of optimized ML models on mobile devices. For instance, MobileNet is a visual model that recognizes one thousand different object classes, which can be designed for efficient execution on mobile and embedded devices.

\section{Example: Deep Network Analyzer for Intelligent Network Management}

The mobile network is fundamentally heterogenous and consists of a variety of different cells such as the macrocell and the small cell. These cells are of different sizes and serve users with diverse traffic demands. For example, the macrocell is usually used to provide radio coverage for a large rural area while a small cell typically serves a urban hot spot area with dense users. In addition, the network conditions of a cellular network can be highly dynamic due to the changing environments and traffic conditions. In this section, we show an example of ML based intelligent network management system for cellular systems. A tool called deep network analyzer (DNA) is developed, which is capable of detecting and diagnosing the anomalous network behaviors at the cell level. The proposed algorithms have been implemented on the Spark platform and have been deployed by Tier-1 operators for intelligent network management. Note that while this network management tool is developed primarily for SQM of 4G system, the proposed ML framework is general and thus can be easily extended to 5G system.

The anomaly detection of DNA relies on a number of datasets including CDR (call detailed record), KQI (key quality indicator) and KPI (key performance indicator). The CDR consists of various information about telephonic calls such as time and duration of the call, source and destination number, the total charged bills, etc. The KQI/KPI datasets contain data files recording the KQI/KPI values of mobile users within a certain time span. The KQIs are indicators that characterize the quality of services (QoSs) of mobile users while KPIs reflect network conditions such as network delay and data packet loss rate. The primary anomalous behavior we aim to uncover is the degradation of network performance.

DNA focuses on the cell-level analytics, in particular cell-level anomaly detection and root cause analysis for wireless cellular network. It is composed of two modules, namely the fingerprint learning module and detection as well as diagnosis modules of abnormal network event, as demonstrated in Figure 3. The fingerprint learning module is responsible for building a database of root cause knowledge from historical data to guide the root cause analysis (RCA), while the anomaly detection (AD) approach is to detect the abnormal behaviors in network that represents performance degradation in user experience. The fingerprint database can also be updated periodically when the new training data is available \cite{yang2017deep}.

\begin{center}
\includegraphics[width=3in]{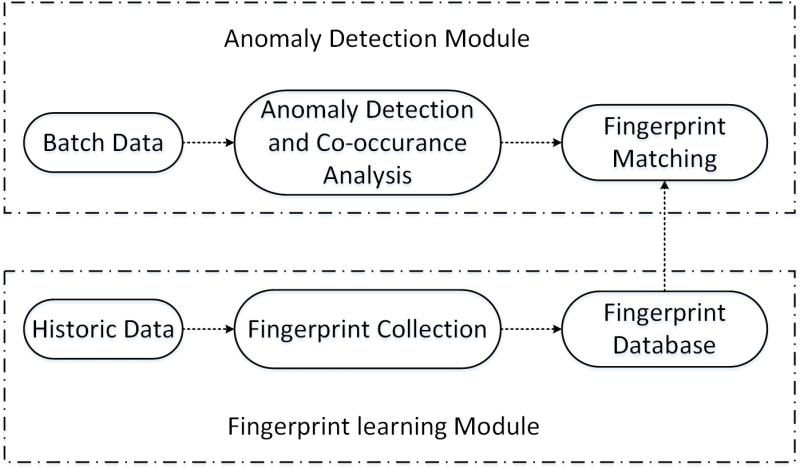}\\
{\ Figure 3.}  The flow diagram of Deep Network Analyzer(DNA)
\end{center}

DNA uses KQIs and KPIs for cell-level network monitoring and analytics. The AD in DNA consists of three stages, namely the data cleaning, anomaly detector and post-processing filtering. During the data cleaning stage, the missing values and extreme values are removed for the training datasets to avoid corrupting the model training in the second stage. Then the datasets is spit into two parts for model training and testing. A statistical univariate anomaly detection algorithm is proposed for effective detection of the outliers in the KQI and KPI datasets. In addition, a set of post-processing filters are introduced to further refine the detect outcomes and remove false alarms in the final stage.

In terms of fingerprint learning module, an approach of rare association rule mining (RARM) is proposed to identify the associations between KQI and KPI. Firstly, the fingerprint knowledge is learned with the aid of RARM method, which uncovers the association rules between KQI and KPIs using a modified FP-Growth algorithm. This modified FP-Growth algorithm can be implemented in a fully distributed manner and thus is well suited for large-scale datasets. Once an anomaly is detected, it will be matched via a K-Nearest Neighbor (KNN) algorithm to the existing signatures in the knowledge database. This tool has been implemented on the open source Spark platform and has been shown to be an effective tool for cell-level network management.

DNA can be viewed as an exemplary fog-intelligence platform for SQM of cellular network. As one example, Spark servers in DNA can either be placed in the cloud or close to the radio network controller (RNC) to collect surveillance data. The former case is suitable for a small to medium-scale wireless network that does not require immediate response to network service outage. In this case, both the ML model building and the inference task can be carried out on the cloud server. The latter case suites better for a large-scale wireless network with tens of millions mobile users in which it is costly and time-consuming to transmit the entire surveillance data to a centralized cloud site. The surveillance data can be stored and processed locally and it allow quick response to abnormal network conditions. In the case of mission-critical and real-time applications such as system outage monitoring and autonomous driving, the server can even be placed to the base station to guarantee real time response and safe operation of the mission-critical systems.

\section{Conclusion}
Future 5G networks are envisioned to support the transmission of massive amount of data from numerous mobile users and IoT devices. However, its management is facing a number of challenges due to its unprecedented scale and complexity. In this paper, we propose fog intelligence as a means to tackle network management problem for future wireless networks. Fog intelligence can be very effective in building distributed machine learning models for intelligent network management while preserving the data privacy as well as reducing the data movement and the processing delays.

\ifCLASSOPTIONcaptionsoff
  \newpage
\fi

\bibliographystyle{IEEEtran}
\bibliography{myreference}

\begin{IEEEbiography}[{\includegraphics[width=1in,height=1.25in,clip,keepaspectratio]{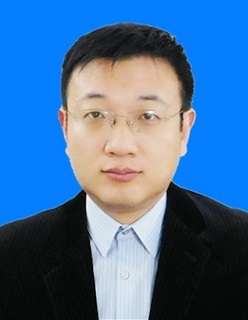}}]{Kai}

\textbf{Yang}(kaiyang@tongji.edu.cn) received the B.Eng. degree from Southeast University, Nanjing, China, the M.S. degree from the National University of Singapore, Singapore, and the Ph.D. degree from Columbia University, New York, NY, USA.

He is a Distinguished Professor with Tongji University, Shanghai, China. He was a Technical Staff Member with Bell Laboratories, Murray Hill, NJ, USA, a Senior Data Scientist with Huawei Technologies, Plano, TX, USA, and a Research Associate with NEC Laboratories America, Princeton, NJ, USA. He has also been an Adjunct Faculty Member with Columbia University since 2011. He holds over 20 patents and has been published extensively in leading IEEE journals and conferences. His current research interests include big data analytics, machine learning, wireless communications, and signal processing.

Dr. Yang was a recipient of the Eliahu Jury Award from Columbia University, the Bell Laboratories Teamwork Award, the Huawei Technology Breakthrough Award, and the Huawei Future Star Award. The products he has developed have been deployed by Tier-1 operators and served billions of users worldwide. He serves as an Editor for the IEEE INTERNET OF THINGS JOURNAL and the IEEE COMMUNICATIONS SURVEYS \& TUTORIALS, and a Guest Editor for the IEEE JOURNAL ON SELECTED AREAS IN COMMUNICATIONS. From 2012 to 2014, he was the Vice-Chair of the IEEE ComSoc Multimedia Communications Technical Committee. In 2017, he founded and served as the Chair of the IEEE TCCN Special Interest Group on AI Embedded Cognitive Networks. He has served as the Demo/Poster Co-Chair of the IEEE INFOCOM, the Symposium Co-Chair of the IEEE GLOBECOM, and the Workshop Co-Chair of the IEEE ICME.
\end{IEEEbiography}

\begin{IEEEbiography}[{\includegraphics[width=1in,height=1.25in,clip,keepaspectratio]{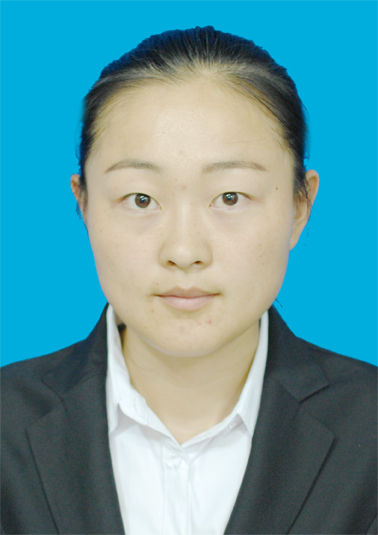}}]{Hui}
\textbf{Ma}(1811437@tongji.edu.cn) was born in Xinjiang, China, in 1991. She received her B.S. degree in 2014 and M.S. degree in 2017 from Jiangnan University, Wuxi, China. She is a Ph.D. candidate in the Department of Computer Science, Tongji University, Shanghai, China. Her current research interests include deep learning and time series forecasting.
\end{IEEEbiography}

\begin{IEEEbiography}[{\includegraphics[width=1in,height=1.25in,clip,keepaspectratio]{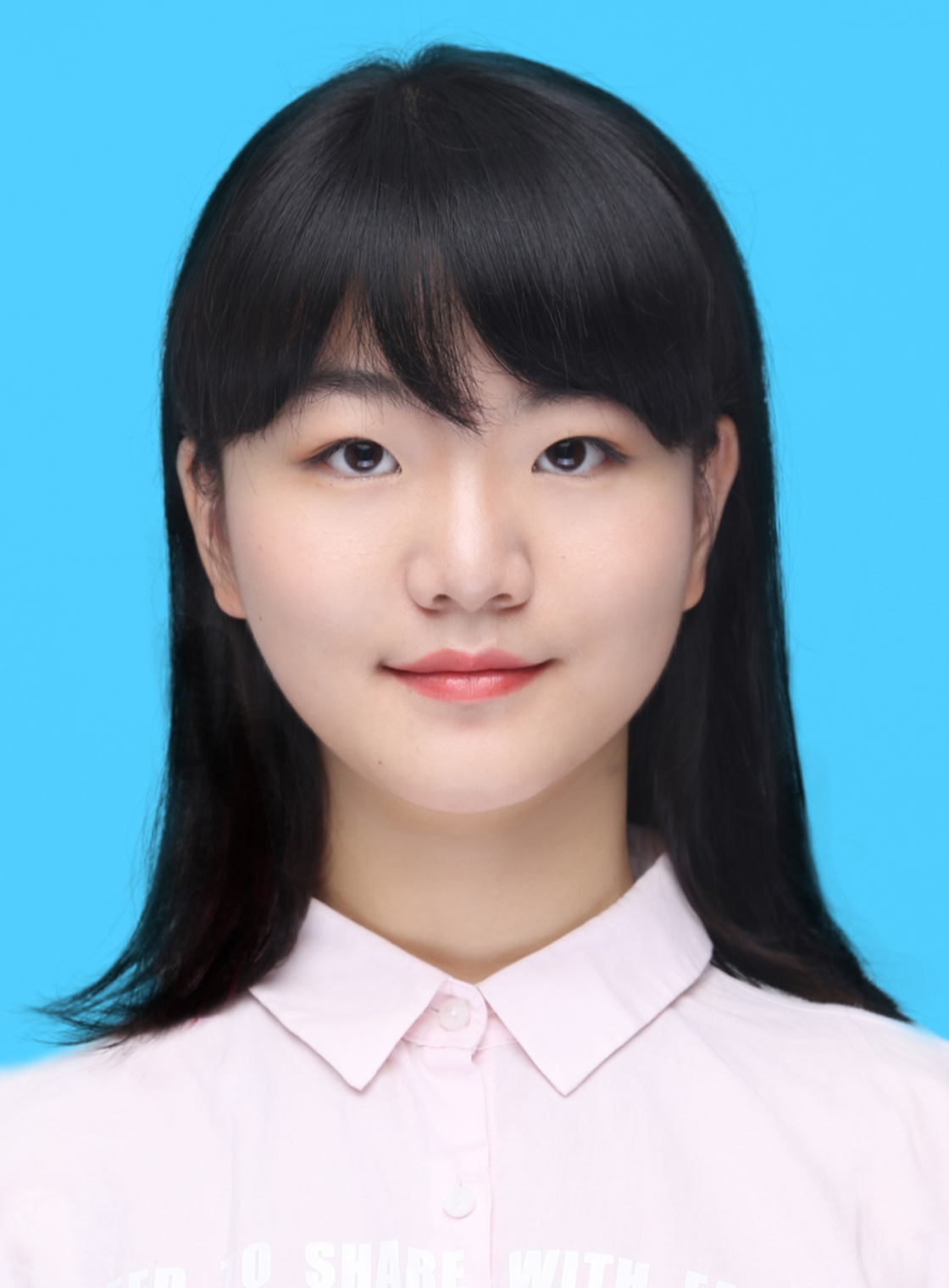}}]{Shaoyu}
\textbf{Dou}(shaoyu@tongji.edu.cn) was born in Gansu, China, in 1996. She received the B.S. degree from Hohai University, Nanjing, China, in 2018. She is currently pursuing the Ph.D. degree in computer science with the Department of Computer Science, Tongji University, Shanghai, China. Her current research interests include big data analytics and machine learning.
\end{IEEEbiography}

\end{document}